\documentstyle[amssymb,aps]{revtex}

\begin{document}

\begin{center}
{\large {\bf Polaron states of electrons over the anisotropic surface of
liquid helium}}

\medskip

Sviatoslav S. Sokolov$^{1,2}$, Ant\^{o}nio Carlos A. Ramos$^{1}$, and Nelson
Studart$^{1}$

\medskip

$^{1}${\small Departamento de F\'{i}sica, Universidade Federal de S\~{a}o
Carlos, 13565-905 S\~{a}o Carlos, S\~{a}o Paulo, Brazil }

$^{2}${\small \ B. I. Verkin Institute for Low Temperature Physics and
Engineering, National Academy of Sciences of Ukraine, 61164 Kharkov, Ukraine 
}

\bigskip\ 

\bigskip Abstract
\end{center}

The energetics and transport properties of the polaron in the anisotropic
surface over liquid helium are investigated. The localization radii and the
energy of the ground and excited states are calculated using the variational
method within the hydrodynamic model of the polaron. In particular, we have
considered maximal anisotropy which corresponds to the system of electrons
in quasi-one-dimensional channels over liquid helium. The polaron binding
energy is found and the temperature for the polaron formation is estimated
below 0.1 K. Solving the hydrodynamic equations for fluid velocities, the
polaron mobilities along and across the channel are determined. The
possibility of experimental observation of polarons by measuring the
frequency of spectroscopic transitions as well the mobility as functions of
the holding electric field is addressed.

\smallskip

\smallskip


\bigskip\ 

\section{INTRODUCTION}

An electron together with its self-induced polarization in a medium forms a
quasiparticle which has been named polaron. Besides its importance as a
standard theoretical model of a fermionic particle coupled to a boson scalar
field, the polaron has been observed in some physical systems. In
particular, there has been great interest in the search of polaron states
for surface electrons levitated over liquid helium whose properties turn out
the system as a good candidate for the formation of a polaron of reduced
dimensionality. Predicted theoretically long time ago,\cite{sh-mon73,mon75}
the surface polaron over liquid helium has been the subject of great amount
of experimental work during the last decades.\cite{experiment,tress,dahm}

Theoretical approaches to investigate the surface polaron over helium are
twofold. One is based on the description of the dimple state (electron plus
the deformation of the isotropic helium surface due to the pressing field)
through the minimization of the total energy functional of the dimple which
leads to a system of coupled equations of motion.\cite
{sh-mon73,mon75,sander75,cheng78} The transport properties are evaluated in
terms of classical hydrodynamical equations since the polaron has a high
effective mass.\cite{sh-mon73,mon75} The other one involves the concept of a
Fr\"{o}hlich-like polaron (a single electron coupled to ripplons).\cite
{ripplonicpolaron} The conductivity was calculated in terms of a force-force
correlation function within the linear response theory.\cite{rippol-transp}
Despite the great difference between the methods, the final results for the
structure of the ripplonic polaron in both approaches show fair qualitative
and quantitative agreement.\cite{review}

In a seminal paper, Shikin and Monarkha\cite{sh-mon73} have determined the
ground state of the electron in an {\it isotropic} surface deformation
through the solution of the Schr\"{o}dinger equation for the electron
trapped in the dimple and the mechanical equilibrium equation using the
Fourier-Bessel transform due to the circular symmetry of the equations. The
profile of the surface deformation was taken in the harmonic approximation
(HA) which allows them to obtain the localization length of the electron
from the Gaussian wave function, i.e. the polaron radius. Later Monarkha\cite
{mon75} using a variational method (VM) was able to obtain the localization
length from the minimization of the polaron energy calculated with a trial
Gaussian wave function. In these works, the influence of external fields and
the thickness of the helium film on the static and dynamical properties of
the polaron was investigated. Marques and Studart\cite{marq-stud89} have
solved in a self-consistent way the Schr\"{o}dinger equation and the
mechanical equilibrium equation obtaining both the electron wave function
and the profile of the dimple. The comparison of the numerical results with
those from the HA and VM shows that the VM provides a more exact description
of the electron wave function than the HA. More recently Farias and Peeters%
\cite{far-peet97} have also used the VM for determining both ground and
excited polaron states taking into account the effect of a positive impurity
charge localized in the substrate that supports the helium film.

Recently, there has been a growing interest in the effects of a corrugated
helium surface in the properties of surface electrons. One motivation is to
make use of suspended helium films\cite{suspended} to increase electron
densities which are limited in the case of bulk helium by a surface
instability and by the impossibility of obtaining high-mobility electrons on
a thin film, in which high densities could be achieved, due to surface
roughness of the substrate. The other one is to confine surface electrons in
one and zero-dimensions as have been realized in semiconductor
heterostructures.

Quasi-one-dimensional (Q1D) electron systems on the surface of liquid helium
have been realized by either geometric or electrostatic mechanisms which
provide the distortion of the helium surface and a confining electric field
holds the surface electrons along the formed liquid channels. Multi-wire
systems have been created using dielectric optical gratings\cite{grating},
substrate wrap by nylon threads\cite{nylon} and metallic gate structures.%
\cite{gate} A single wire was also performed using a sharply bent polymer
film\cite{kirichek} and metallic strips on printed circuit board.\cite
{valkering} In such a Q1D system, the electron motion is restricted, in
addition to the quantum well due to the holding electric in the direction
normal to the liquid surface, by a lateral confinement and can be modelled
in first order of approximation by a harmonic potential. In such conditions,
the formation of an {\it asymmetric} polaron may become also possible.

In this paper, we address the question of polaron properties by using the
hydrodynamic approach in the case of an anisotropic potential $U(x,y)$ due
to surface corrugation and in particular for the Q1D electron system
described by a parabolic confinement in the $y$ direction. The general
formalism can be applied for both circularly symmetric and asymmetric
polaron states. We consider the properties of both ground and excited
polaron states and the anisotropic transport properties of the polaron are
investigated for surface electrons on He$^{4}$ and He$^{3}$.\cite{kono95}

The paper is organized as follows. The general formalism and the main
relations are described in Sec. II. In Sec. III we analyze the properties of
the ground and excited polaron states. In Sec. IV we investigate the polaron
mobility when a driving electric field is applied parallel to the liquid
surface. In Sec. V we summarize our main results.

\section{THEORETICAL FORMALISM}

Electrons above the free surface ($z=0$) of a helium film with thickness $d$
is prevented to penetrate into the liquid because of a high potential
barrier ($\sim 1$ eV) at the liquid-vapor interface. If an electric field $%
E_{\perp }$ is applied in the $z$ direction, the electrons are trapped in a
quantum well due to image forces coming from the liquid helium and substrate
and determined by the potential $V(z)=eE_{\perp }z-\Lambda _{0}/z-\Lambda
_{1}/(z+d)$ where $\Lambda _{0}=e^{2}(\epsilon _{He}-1)/4(\epsilon _{He}+1),$
$\Lambda _{1}=e^{2}\epsilon _{He}(\epsilon _{s}-\epsilon _{He})/(1+\epsilon
_{He})^{2}(\epsilon _{He}+\epsilon _{s})$ with $\epsilon _{He}$ and $%
\epsilon _{s}$ the dielectric constants of helium and substrate
respectively. If the barrier height is approximated as infinity, the
condition $\Psi (x,y,z=0)=0$ for the electron wave function must be
fulfilled for a flat surface. The situation changes drastically if we take
into account the surface deformation $\xi (x,y)$. Now the boundary condition
for $\Psi (x,y,z)$ has to be imposed at $z=\xi (x,y)$. Shikin and Monarkha%
\cite{shi-mon74} show that the transformation to a new variable $z^{\prime
}=z-\xi (x,y)$ allows to avoid the perturbation in the boundary condition
for $\Psi (x,y,z)$ leading, however, to modifications in the Schr\"{o}dinger
equation. For $\xi (0,0)\ll d$ and $\langle z\rangle \ll d$, where $\langle
z\rangle $ is the mean electron distance from the surface, these
modifications result, in particular, in the dependence of $V(z)$ on an
effective holding electric field given by $E_{\perp }^{*}=E_{\perp }+\Lambda
/ed^{2}$ with $\Lambda =e^{2}(\varepsilon _{s}-1)/4(\varepsilon _{s}+1)$
where we take $\varepsilon _{He}\simeq 1$, and the appearance of an
additional term $eE_{\perp }^{*}\xi (x,y)$.

The electron motion along the plane ($x,y$) in the presence of a magnetic
field $B$ in the $z$ direction is described by the Schr\"{o}dinger equation

\begin{equation}
\frac{1}{2m}[(\widehat{p}_{x}+\frac{eB}{2c}y)^{2}+(\widehat{p}_{y}-\frac{eB}{%
2c}x)^{2}]\psi (x,y)+[eE_{\perp }^{*}\xi (x,y)+U(x,y)]\psi (x,y)=\varepsilon
\psi (x,y).  \label{2.1}
\end{equation}
Here $\widehat{p}_{x}$ and $\widehat{p}_{y}$ are the $x$ and $y$ components
of the momentum operator and we have chosen the symmetric gauge of the
vector potential $\vec{A}=(-By/2,Bx/2,0)$. Note that due to the explicit ($%
x,y$)-dependence of the potential term in Eq. (\ref{2.1}), $\hat{p}_{x}$ and 
$\hat{p}_{y}$ are not conserved. The confinement potential $U(x,y)$ will be
considered as a general anisotropic parabolic well given by 
\begin{equation}
U(x,y)=\frac{m\omega _{0}^{2}}{2}(\alpha x^{2}+y^{2}).  \label{2.2}
\end{equation}
where $\alpha $ is the anisotropic parameter. For $\alpha =1,$ we restore
the circular symmetry and $\alpha =0$ corresponds the case of a Q1D electron
considered in Refs. \cite{kov-mon86,sok-hai-stud95}. Then Eq. (\ref{2.1})
can be rewritten as 
\begin{equation}
-\frac{\hbar ^{2}}{2m}{\bf \nabla }^{2}\psi -\frac{\hbar \omega _{c}}{2}%
\widehat{L}_{z}\psi +[\frac{m}{8}\left( \omega _{x}^{2}x^{2}+\omega
_{y}^{2}y^{2}\right) +eE_{\perp }^{*}\xi ]\psi =\varepsilon \psi ,
\label{2.3}
\end{equation}
where ${\bf \nabla }$ is the 2D-Laplacian, $\omega _{x}^{2}=\omega
_{c}^{2}+4\alpha \omega _{0}^{2}$, $\omega _{y}^{2}=\omega _{c}^{2}+4\omega
_{0}^{2}$ with $\omega _{c}=eB/mc$ the cyclotron frequency, and $\widehat{L}%
_{z}$ is the angular momentum operator along $z$ which is also not conserved
for $\alpha \neq 1$ when the axial symmetry is lost. As a consequence, the
second term in Eq. (2) does not contribute to energy eigenvalues $%
\varepsilon $ if the $\psi (x,y)$ is taken real. If we assume that $\psi $
is real and vanishes at infinity the electron energy can be calculated from 
\begin{equation}
\varepsilon =\int \int \left[ \frac{\hbar ^{2}}{2m}\left( {\bf \nabla }\psi
\right) ^{2}+\frac{m}{8}\left( \omega _{x}^{2}x^{2}+\omega
_{y}^{2}y^{2}\right) \psi ^{2}+eE_{\perp }^{*}\xi (x,y)\psi ^{2}\right] dxdy{%
.}  \label{2.4}
\end{equation}
The total energy of the complex ``electron + dimple'' can be defined as 
\begin{equation}
W=\varepsilon +\frac{\sigma }{2}\int \int \left[ \left( {\bf \nabla }\xi
\right) ^{2}+k_{c}^{2}\xi ^{2}\right] dxdy{,}  \label{2.5}
\end{equation}
where $k_{c}^{2}=\rho g^{\prime }/\sigma $ is the capillary constant, $%
g^{\prime }=g(1+3f/\rho gd^{4})$, $f$ is the van der Waals coupling constant
of the helium to the substrate, $\sigma $ and $\rho $ are the surface
tension coefficient and the mass density of helium, respectively, and $g$ is
the gravitational acceleration. As is seen from Eq. (\ref{2.5}) the
inequality $W>\varepsilon $ is always satisfied. Minimizing the Eq. (\ref
{2.5}), we obtain the mechanical equilibrium equation 
\begin{equation}
\sigma (\nabla ^{2}\xi -k_{c}^{2}\xi )=eE_{\perp }^{*}\psi ^{2}{.}
\label{2.6}
\end{equation}
Note that the quantity of $eE_{\perp }^{*}\psi ^{2}$ plays the role of an
electron pressure on the liquid surface.

As is seen from Eqs. (\ref{2.3})-(\ref{2.5}), the asymmetry of the electron
motion in $x$ and $y$ directions for $\omega _{x}\neq \omega _{y}$ makes
inappropriate the use of the Fourier-Bessel transform in polar coordinates
as in previous works. Here we use the 2D Fourier transform for $\xi (x,y)$
as 
\begin{equation}
\xi (x,y)=\sum_{{\bf k}}\xi _{{\bf k}}e^{i(k_{x}x+k_{y}y)};\hspace{0.5cm}\xi
_{{\bf k}}=\frac{1}{S}\int \xi (x,y)e^{-i(k_{x}x+k_{y}y)}dxdy{,}  \label{2.7}
\end{equation}
and the similar transform for $\psi $. Here ${\bf k}$ is the 2D wave vector
and $S$ is the surface area. Using Eq. (\ref{2.7}) one can easily obtain the
following expression which connects the Fourier transforms of $\xi (x,y)$
and $\psi ^{2}(x,y)$: 
\begin{equation}
\xi _{{\bf k}}=-\frac{eE_{\perp }^{*}}{\sigma (k^{2}+k_{c}^{2})}\left[ \psi
^{2}(x,y)\right] _{{\bf k}}{.}  \label{2.8}
\end{equation}
Equations (\ref{2.7}) and (\ref{2.8}) can be used to eliminate $\xi (x,y)$
from Eqs. (\ref{2.4}) and (\ref{2.5}) and obtain the total energy of the
polaron in terms only of the electron wave function and its Fourier
transform as 
\begin{eqnarray}
W &=&\int \int \left[ \frac{\hbar ^{2}}{2m}\left( {\bf \nabla }\psi \right)
^{2}+\frac{m}{8}\left( \omega _{x}^{2}x^{2}+\omega _{y}^{2}y^{2}\right) \psi
^{2}\right] dxdy  \nonumber \\
&&-\frac{(eE_{\bot }^{*})^{2}}{2\sigma }\sum_{{\bf k}}\frac{\left[ \psi
^{2}(x,y)\right] _{{\bf k}}\left[ \psi ^{2}(x,y)\right] _{-{\bf k}}}{%
k^{2}+k_{c}^{2}}{.}  \label{2.9}
\end{eqnarray}
The electron energy $\varepsilon $ has the same expression as $W$ except by
the absence of the factor $2$ in the last term of Eq. (\ref{2.9}). For the
sake of completeness, we can rewrite the resulting the Schr\"{o}dinger
equation after removing $\xi (x,y)$ from Eq. (\ref{2.3}), as: 
\[
-\frac{\hbar ^{2}}{2m}{\bf \nabla }^{2}\psi -\frac{\hbar \omega _{c}}{2}%
\widehat{L}_{z}\psi +\frac{m}{8}\left( \omega _{x}^{2}x^{2}+\omega
_{y}^{2}y^{2}\right) \psi 
\]
\begin{equation}
-\frac{(eE_{\bot }^{*})^{2}}{\sigma }\left( \sum_{{\bf k}}\frac{[\psi
^{2}(x,y)]_{{\bf k}}}{k^{2}+k_{c}^{2}}e^{i(k_{x}x+k_{y}y)}\right) \psi
=\varepsilon \psi {.}  \label{2.10}
\end{equation}

Equation (\ref{2.10}) is quite general and can be solved self-consistently
and the results are used to evaluate the total polaron energy $W$ and
surface profile $\xi (x,y)$. However this procedure is totally numerical and
cumbersome even in the symmetric case $\alpha =1$.\cite{marq-stud89} We
prefer to make reasonable guesses about the structure of $\psi (x,y)$ and
obtain analytical results for the energetics and transport properties of the
asymmetric polaron.

We also hereafter limit ourselves to the most studied case of the Q1D system
model\cite{kov-mon86,sok-hai-stud95}, where the surface electrons are
confined by a lateral potential well corresponding the completely
anisotropic limit, $\alpha =0$, i.e. $U(y)=m\omega _{0}^{2}y^{2}/2$ with the
characteristic frequency defined as $\omega _{0}=\sqrt{eE_{\perp }^{*}/mR}$,
where $R$ ($\sim 10^{-4}-10^{-3}$ cm) is the curvature radius of the liquid
in the channel. In the absence of a magnetic field the spectrum for the free
electron motion is given by $E(k_{x},n)=\hslash ^{2}k_{x}^{2}/2m+\hslash
\omega _{0}(n+1/2)$ and the electron localization in the $y$ direction
estimated by the parameter $L_{0}=(\hslash /m\omega _{0})^{1/2}\ll R$ and
for this reason, curvature effects play no significant role because $%
E_{\perp }^{*}$ pushes the electron to the bottom of the channel. This
parameter somewhat changes in the presence of a magnetic field.\cite
{sok-stud95} Typical values of $L_{0}$ are of the order of $10^{-6}$ cm\cite
{kov-mon86,sok-hai-stud95} and the condition $L_{0}\ll R$ is satisfied with
great accuracy. Obviously for the polaron state the localization parameter
along the $y$ direction will be significantly modified.

\section{POLARON ENERGETICS}

\subsection{Ground-state}

Based on the general structure of Eq. (\ref{2.10}) we choose the trial
function to describe the ground state of the polaron as 
\begin{equation}
\psi _{0}(x,y)=\frac{1}{\pi ^{1/2}(\ell _{x}\ell _{y})^{1/2}}\exp \left[ -%
\frac{1}{2}\left( \frac{x^{2}}{\ell _{x}^{2}}+\frac{y^{2}}{\ell _{y}^{2}}%
\right) \right] {,}  \label{3.1}
\end{equation}
where $\ell _{x}$ and $\ell _{y}$ are the electron localization lengths in $%
x $ and $y$ directions, respectively. Substituting Eq. (\ref{3.1}) into Eqs.
(\ref{2.6})-(\ref{2.9}), one obtains the surface deformation $\xi _{0}(x,y)$%
, and the total energy $W_{0}$ of the polaron ground state as

\begin{equation}
\xi _{0}(x,y)=-\frac{eE_{\perp }^{*}}{4\pi ^{2}\sigma }\int dk_{x}\int dk_{y}%
\frac{\exp \left[ -\left( k_{x}^{2}\ell _{x}^{2}+k_{y}^{2}\ell
_{y}^{2}\right) /4\right] \cos (k_{x}x)\cos (k_{y}y)}{%
k_{x}^{2}+k_{y}^{2}+k_{c}^{2}}  \label{3.2}
\end{equation}
and 
\begin{eqnarray}
W_{0} &=&-\frac{(eE_{\perp }^{*})^{2}}{8\pi ^{2}\sigma }\int dk_{x}\int
dk_{y}\frac{\exp \left[ -\left( k_{x}^{2}\ell _{x}^{2}+k_{y}^{2}\ell
_{y}^{2}\right) /2\right] }{k_{x}^{2}+k_{y}^{2}+k_{c}^{2}}+\frac{\hbar ^{2}}{%
4m}\left( \frac{1}{\ell _{x}^{2}}+\frac{1}{\ell _{y}^{2}}\right)  \nonumber
\\
&&+\frac{m}{16}\left[ \omega _{c}^{2}\ell _{x}^{2}+\left( \omega
_{c}^{2}+4\omega _{0}^{2}\right) \ell _{y}^{2}\right] {.}  \label{3.3}
\end{eqnarray}
The ground state energy $\varepsilon ^{(0)}$ of electron has the same form
except by the coefficient of the integral which in this case is $(eE_{\perp
}^{*})^{2}/4\pi ^{2}\sigma .$ In the limit $k_{c}^{2}(\ell _{x}^{2}+\ell
_{y}^{2})\ll 1$, the integral in Eq. (\ref{3.3}) can be evaluated
analytically resulting in 
\begin{equation}
W_{0}\simeq -\frac{(eE_{\perp }^{*})^{2}}{4\pi \sigma }\ln \left[ \frac{2%
\sqrt{2}}{\sqrt{\gamma }k_{c}(\ell _{x}+\ell _{y})}\right] +\frac{\hbar ^{2}%
}{4m}\left( \frac{1}{\ell _{x}^{2}}+\frac{1}{\ell _{y}^{2}}\right) +\frac{m}{%
16}\left[ \omega _{c}^{2}\ell _{x}^{2}+\left( \omega _{c}^{2}+4\omega
_{0}^{2}\right) \ell _{y}^{2}\right] ,  \label{3.4}
\end{equation}
where $\gamma =\exp C$, and $C=0.5772...$ is the Euler-Mascheroni constant.
The value of $\varepsilon ^{(0)}$ is near the same obtaining the coefficient 
$2$ instead of $4$ in the denominator of first term of Eq. (\ref{3.4}).

The localization lengths $\ell _{x}$ and $\ell _{y}$, which appear in Eqs. (%
\ref{3.1})-(\ref{3.4}), have been evaluated by a few methods. In particular,
the HA similar to that used in Ref. \cite{sh-mon73} can be applied and is
based on the following approximate expression for the surface deformation

\begin{equation}
\xi _{0}(x,y)\simeq {\xi (0,0)+}\frac{1}{2}\left[ \xi _{xx}^{^{\prime \prime
}}(0,0)x^{2}+\xi _{yy}^{^{\prime \prime }}(0,0)y^{2}\right] .  \label{3.5}
\end{equation}
The expansion is similar to that of the potential energy near its minimum
value in the 2D oscillatory problem and reduces the problem of the electron
motion in the dimple to the motion in a parabolic confinement potential, by
defining $\ell _{x}$ and $\ell _{y}$ as localizations parameters of the 2D
harmonic oscillator. We prefer however to use the VM which allows to obtain $%
\ell _{x}$ and $\ell _{y}$ from the minimization conditions of the energy $%
W_{0}$, i.e. $\partial W_{0}/\partial \ell _{x}=\partial W_{0}/\partial \ell
_{y}=0$ of the minimum. Note that in the VM the key point for consideration
is the expression for the polaron energy, given by the Eq. (\ref{3.4}),
whereas the structure of the dimple potential is less important and is taken
into account through Eq. (\ref{2.6}). From the experimental point of view,
using the VM seems more convenient because the characteristic value of the
helium surface depression in the centre of the dimple is about $10^{-8}$ cm
for $E_{\perp }^{*}\sim 3$ kV/cm which is impossible to detect directly.
Hereafter this estimative is based on actual holding fields in experiments
on electrons along Q1D channels on bulk helium, where the effects of the
film thickness are neglected.\cite{grating,kirichek} On the other hand, as
will see, the energy gap between the ground and excited polaron states,
calculated using the VM, may in principle be accessible experimentally.

In the VM, $\ell _{x}$ and $\ell _{y}$ can be find from the roots of the
system of equations

\begin{equation}
\frac{1}{\ell _{x}^{4}}-\frac{1}{L_{F}^{2}\ell _{x}(\ell _{x}+\ell _{y})}-%
\frac{1}{L_{B}^{4}}=0\;\text{and}\;\frac{1}{\ell _{y}^{4}}-\frac{1}{%
L_{F}^{2}\ell _{y}(\ell _{x}+\ell _{y})}-\frac{1}{L_{0}^{4}}-\frac{1}{%
L_{B}^{4}}=0{,}  \label{3.6}
\end{equation}
where $L_{F}^{2}=2\pi \sigma \hslash ^{2}/m(eE_{\perp }^{*})^{2}$ and $%
L_{B}^{2}=2\hslash /m\omega _{c}$. Eqs. (\ref{3.6}) have been solved
numerically. The results are presented in Fig. 1 as a function of the
holding field for some values of the magnetic field and for $^{4}$He as the
liquid substrate. The analytical solution of Eqs. (\ref{3.6}) is possible in
some limiting cases. For $B=0$ and very high holding fields $E_{\perp
}^{*}\gg 4$ kV/cm satisfying the condition $L_{F}\ll L_{0}$, one obtains $%
\ell _{x}\simeq \ell _{y}\simeq \sqrt{2}L_{F}$ which is the localization
parameter for the symmetric polaron where the effects of the lateral
confinement along $y$ are negligible. For holding fields in the range $%
1<E_{\perp }^{*}<3$ kV/cm, in the opposite limit $L_{F}\gg L_{0}$, one has $%
\ell _{x}\simeq L_{F}$ and $\ell _{y}\simeq L_{0}$ which correspond to the
numerical results shown in Fig. 1. Hence the localization length $\ell _{y}$
is almost the same as that for the electron moving freely along the Q1D
channel. Numerical estimates at $B=0$ are $\ell _{x}\sim 10^{-5}$ cm and $%
\ell _{y}\sim 10^{-6}$ cm for $1<E_{\perp }^{*}<3$ kV/cm. These values are
significantly smaller than the curvature radius $R$. It means that not only
the condition $\ell _{y}\ll R$, which supports the validity of our Q1D
confinement model, is fulfilled but also the condition $\ell _{x}\ll R$ is
satisfied. The application of the magnetic field leads to a decrease of
localization lengths in comparison with those for $B=0$. For high $B$ ($%
\omega _{c}\gg \omega _{0}$), one obtains from Eqs. (\ref{3.6}) $\ell
_{x}\simeq \ell _{y}\simeq \sqrt{2}L_{F}$ for $L_{F}\ll L_{B}$ and $\ell
_{x}\simeq \ell _{y}\simeq L_{B}$ for $L_{F}\gg L_{B}.$ We point out that $%
\ell _{x}$ and $\ell _{y}$, calculated using the HA, give different
asymptotic values: $\ell _{x}\simeq \ell _{y}\simeq L_{F}$ for $L_{F}\ll
L_{0},L_{B}$ and for $B=0$, the results are $\ell _{x}\simeq L_{F}/\sqrt{2}$
and $\ell _{y}\simeq L_{0}$ at $L_{F}\gg L_{0}$. For high magnetic fields
and $L_{F}\gg L_{B},$ one has $\ell _{x}\simeq \ell _{y}\simeq L_{B}$. It is
interesting also to note that if the VM is used for the energy of the
electron trapped in the dimple $\varepsilon ^{(0)}$, instead of $W_{0}$, the
results are the same as those obtained in HA. One can conclude the solutions
of Eqs. (\ref{3.6}) yielding $\ell _{x}=\ell _{y}$, which is the case of a
symmetric polaron, appear either in the limit of very high holding field
(small $L_{F}$) or for high magnetic fields ($\omega _{c}\gg \omega _{0},$ $%
L_{F}\gg L_{B})$.

Using the localization lengths from Eq. (\ref{3.6}), we depicted in Fig. 2
the polaron energy as a function of the holding field for some values of the
magnetic field. For $\ell _{x}\simeq L_{F}$ and $\ell _{y}\simeq L_{0}$ and $%
B=0$, Eq. (\ref{3.4}) is rewritten as

\begin{equation}
W_{0}\simeq -\frac{(eE_{\perp }^{*})^{2}}{4\pi \sigma }\ln \frac{2\sqrt{2}}{%
\sqrt{\gamma }k_{c}L_{F}}+\frac{\hbar ^{2}}{4mL_{F}^{2}}{+}\frac{\hbar
\omega _{0}}{2}.  \label{3.7}
\end{equation}
Defining the binding energy $E_{b}$ as the energy of the polaron state minus
the electron energy $\hbar \omega _{0}/2$ of the free motion in the lowest
subband of the lateral potential, one obtains 
\begin{equation}
E_{b}\simeq -\frac{(eE_{\perp }^{*})^{2}}{4\pi \sigma }\ln \frac{2\sqrt{2}}{%
\sqrt{\gamma }k_{c}L_{F}}+\frac{\hbar ^{2}}{4mL_{F}^{2}}{.}  \label{3.8}
\end{equation}
Note that we have considered very low temperatures $T\ll \hbar \omega _{0}$
where only the lowest subband $n=0$ is occupied (for the classical SE system 
$k_{x}\sim \sqrt{mT}/\hbar $). The polaron state is preferable energetically
at $T<|E_{b}|$, otherwise thermal motion can liberate electron from the
dimple. We estimate $E_{b}\simeq -0.03$ K and $E_{b}\simeq -0.3$ K for
holding electric fields of $1$ V/cm and $3$ kV/cm, respectively, in the case
of a $^{4}$He substrate. One can conclude that the energetic conditions for
the formation of the polaron in the Q1D electron system on the liquid helium
surface are almost the same as in the case of the 2D symmetric polaron.

For magnetic fields satisfying the conditions $\omega _{c}\gg \omega _{0},$ $%
L_{F}\ll L_{B}$, the binding energy, defined by extracting the cyclotron
energy $\hbar \omega _{c}/2$ of the lowest Landau level from $W_{0}$ in Eq.(%
\ref{3.4}) can be written as

\begin{mathletters}
\begin{equation}
E_{b}\simeq -\frac{(eE_{\perp }^{*})^{2}}{4\pi \sigma }\ln \frac{\sqrt{2}}{%
\sqrt{\gamma }k_{c}L_{B}},  \label{3.9}
\end{equation}
which agrees with the asymptotic value of $E_{b}$ for the symmetric polaron $%
(\omega _{0}=0)$ for very high $B$. For $B=5.5$ T $(\omega _{c}=10^{12}$ s$%
^{-1}$), we estimate $L_{B}=1.52\times 10^{-6}$ cm and $E_{b}\simeq -0.36$ K
for $E_{\perp }^{*}=3$ kV/cm, which is very close to that for $B=0$.

We have also estimated the profile of the surface from Eq. (\ref{3.2}). For
large distances such that $k_{c}x$ and $k_{c}y\gg 1$, $\xi (x,y)$ decreases
exponentially as a function of the distance $r=\sqrt{x^{2}+y^{2}}$ in the
same manner as for symmetric polaron.\cite{marq-stud89} The value of the
dimple depth in its centre can be written as

\end{mathletters}
\begin{equation}
\xi _{0}(0,0)\simeq -\frac{eE_{\perp }^{*}}{2\pi \sigma }\ln \frac{4}{\sqrt{%
\gamma }k_{c}(\ell _{x}+\ell _{y})}.  \label{3.10}
\end{equation}
For $B=0$ and $\ell _{x}\simeq L_{F},$ $\ell _{y}\simeq L_{0}$, we obtain $%
\xi _{0}(0,0)\simeq -(eE_{\perp }^{*}/2\pi \sigma )\ln (4/\sqrt{\gamma }%
k_{c}L_{F})\simeq -1.9\times 10^{-8}$ cm for $E_{\perp }^{*}=3$ kV/cm if the
liquid substrate is $^{4}$He. For high $B$, where $L_{F}\gg L_{B}$ and $%
\omega _{c}\gg \omega _{0}$, one has $\xi _{0}(0,0)\simeq -(eE_{\perp
}^{*}/2\pi \sigma )\ln (2/\sqrt{\gamma }k_{c}L_{B})\simeq -2.2\times 10^{-8}$
cm.

It is interesting how to recover from our results those for the symmetric
polaron where $\omega _{0}=0$. In this case $x$ and $y$ directions are
equivalent and the $z$ component of the angular moment conserves. Due to
this one can replace the operator $\widehat{L}_{z}$ of angular moment in
Eqs. (\ref{2.3}) and (\ref{2.10}) by its eigenvalue $l_{z}$ which leads to
the new contribution $\hbar \omega _{c}l_{z}/2$ to the electron energy $%
\varepsilon \ $in the case of non-zero magnetic field. We should put $%
l_{z}=0 $ in the ground state of the polaron with lowest angular momentum.
Moreover this contribution is zero for real electron wave functions in the
polaron state even for nonzero angular momentum. If $\omega _{0}=0$, the
solution of Eqs. (\ref{3.6}) is $\ell =\ell _{x}=\ell _{y}$ with $1/\ell
^{2}=1/4L_{F}^{2}+\sqrt{1/16L_{F}^{4}+1/L_{B}^{2}}$ and the electron wave
function for the ground state is written as 
\[
\psi _{0}(x,y)=\frac{1}{\pi ^{1/2}\ell }e^{-r^{2}/2\ell ^{2}} 
\]
which leads to the same results for $W_{0},$ $\xi _{0}(r)$ as previously
found in the case of a symmetric polaron.\cite{sh-mon73,mon75,marq-stud89}
In particular, the polaron energy is 
\[
W_{0}\simeq -\frac{(eE_{\perp })^{2}}{4\pi \sigma }\left[ \ln \frac{1}{\sqrt{%
\gamma }k_{c}L_{F}}-1\right] {.} 
\]
One can easily also obtain that the second-order derivative of $\xi (r)$ at
the centre of the dimple $\left[ \xi _{rr}^{\prime \prime }(0,0)\right]
_{0}=eE_{\perp }^{*}\ /2\pi \sigma \ell ^{2}$ which shows the existence of
the minimum at $r=0$.

\subsection{Excited states}

We must choose a trial excited-state wave function orthogonal to ground
state wave function given by Eq. (\ref{3.1}), which as we have seen is the
same as wave function for the 2D asymmetric harmonic oscillator in Cartesian
coordinates. Hence it is natural to propose the wave function of excited
states of the 2D harmonic oscillator as the trial functions for the excited
polaron states: 
\begin{equation}
\psi _{10}(x,y)=\frac{\sqrt{2}x}{\pi ^{1/2}(\delta _{x}^{3}\delta _{y})^{1/2}%
}\exp \left[ -\frac{1}{2}\left( \frac{x^{2}}{\delta _{x}^{2}}+\frac{y^{2}}{%
\delta _{y}^{2}}\right) \right] {.}  \label{3.11}
\end{equation}
Evidently the another excited state $\left| 0,1\right\rangle $ can be
considered with wave function $\psi _{01}(x,y).$ The wave function $\psi
_{01}$ has the same form as $\psi _{10}$ by replacing $x$ by $y$ in Eq. (\ref
{3.11}). We find however that the state $\left| 1,0\right\rangle $ has
energy smaller than $\left| 0,1\right\rangle $. For this reason the state $%
\left| 1,0\right\rangle $ should be considered the first excited polaron
state and we are looking for results for this state. The results for the
state $\left| 0,1\right\rangle $ can be easily obtained in a straightforward
way.

Following the same procedure as for the ground state we arrive to following
expressions of the polaron energy and the profile of the dimple in excited
state: 
\begin{eqnarray}
W_{10} &=&-\frac{(eE_{\perp })^{2}}{8\pi ^{2}\sigma }\int dk_{x}\int dk_{y}%
\frac{\left( 1-k_{x}^{2}\delta _{x}^{2}/2\right) ^{2}\exp \left[ -\left(
k_{x}^{2}\delta _{x}^{2}+k_{y}^{2}\delta _{y}^{2}\right) /2\right] }{%
k_{x}^{2}+k_{y}^{2}+k_{c}^{2}}+\frac{\hbar ^{2}}{4m}\left( \frac{3}{\delta
_{x}^{2}}+\frac{1}{\delta _{y}^{2}}\right)  \nonumber \\
&&+\frac{m}{16}\left[ 3\omega _{c}^{2}\delta _{x}^{2}+\left( \omega
_{c}^{2}+4\omega _{0}^{2}\right) \delta _{y}^{2}\right]  \label{3.12}
\end{eqnarray}
and 
\begin{equation}
\xi _{10}(x,y)=-\frac{eE_{\perp }}{4\pi ^{2}\sigma }\int dk_{x}\int dk_{y}%
\frac{\left( 1-k_{x}^{2}\delta _{x}^{2}/2\right) \exp \left[ -\left(
k_{x}^{2}\delta _{x}^{2}+k_{y}^{2}\delta _{y}^{2}\right) /4\right] \cos
(k_{x}x)\cos (k_{y}y)}{k_{x}^{2}+k_{y}^{2}+k_{c}^{2}}.  \label{3.13}
\end{equation}
As before the electron energy $\varepsilon ^{(10)}$ has the coefficient $%
-(eE_{\perp }^{*})^{2}/4\pi ^{2}\sigma $ in first term in Eq. (\ref{3.12}).

The $x$ dependence of $\psi _{10}(x,y)$ contributes strongly to decrease the
electronic pressure on the liquid surface at $x=0$, at the centre of the
dimple. According to Eq. (\ref{2.6}), the pressure is proportional to $\psi
_{10}^{2}(x,y).$ As a consequence, in the limit $k_{x}^{2}(\delta
_{x}^{2}+\delta _{y}^{2})\ll 1$, the second-order partial derivative $[\xi
_{xx}^{\prime \prime }]_{10}(0,0)$ calculated by Eq. (\ref{3.13}) $[\xi
_{xx}^{\prime \prime }]_{10}(0,0)=-[\xi _{yy}^{\prime \prime
}]_{10}(0,0)\simeq -eE_{\perp }^{*}/[\pi \sigma (\delta _{x}+\delta
_{y})^{2}].$ This result is a direct consequence of the structure of the
trial wave function $\psi _{10}$\smallskip $(x,y)$ and means that the
function $\xi _{10}(x,y)$ has a maximum at $x=0$ for fixed $y$ whereas a
minimum at $y=0$ at fixed $x$ and one can not conclude definitely about the
nature of the dimple potential as a function of two variables $x$ and $y$ at 
$x=y=0$. In such condition the use of the HA to excited states is
inappropriate because it is based on the expansion given by Eq. (\ref{3.5})
near the {\it minimum} of the dimple potential at $x=y=0.$ For this reason
the VM is the only consistent way to calculate the polaron properties in the
excited state.

As before the localization lengths $\delta _{x}$ and $\delta _{y}$ can be
obtained, in VM, from the conditions of a minimum of energy $W_{10}$ given
by Eq. (\ref{3.12}). For $k_{c}^{2}(\delta _{x}^{2}+\delta _{y}^{2})\ll 1$,
we obtain

\begin{eqnarray}
W_{10} &\simeq &-\frac{(eE_{\perp }^{*})^{2}}{4\pi \sigma }\left[ \ln \frac{2%
\sqrt{2}}{\sqrt{\gamma }k_{c}(\delta _{x}+\delta _{y})}-\frac{\delta
_{x}(2\delta _{x}+3\delta _{y})}{4(\delta _{x}+\delta _{y})^{2}}\right] +%
\frac{\hbar ^{2}}{4m}\left( \frac{3}{\delta _{x}^{2}}+\frac{1}{\delta
_{y}^{2}}\right)  \nonumber \\
&&+\frac{m}{16}\left[ 3\omega _{c}^{2}\delta _{x}^{2}+\left( \omega
_{0}^{2}+4\omega _{c}^{2}\delta _{y}^{2}\right) \right] {.}  \label{3.14}
\end{eqnarray}
Imposing that $\partial W_{10}/\partial x=\partial W_{10}/\partial y$%
\smallskip $=0$ we arrive to the system of equations

\begin{eqnarray}
\frac{3}{\delta _{x}^{4}}-\frac{1}{L_{F}^{2}\delta _{x}(\delta _{x}+\delta
_{y})}\left[ 1+\frac{\delta _{y}(\delta _{x}+3\delta _{y})}{4(\delta
_{x}+\delta _{y})^{2}}\right] -\frac{3}{L_{B}^{4}} &=&0  \nonumber \\
\frac{1}{\delta _{y}^{4}}-\frac{1}{L_{F}^{2}\delta _{y}(\delta _{x}+\delta
_{y})}\left[ 1-\frac{\delta _{x}(\delta _{x}+3\delta _{y})}{4(\delta
_{x}+\delta _{y})^{2}}\right] -\frac{1}{L_{0}^{4}}-\frac{1}{L_{B}^{4}} &=&0{.%
}  \label{3.15}
\end{eqnarray}
In the limit of very high holding fields where $L_{F}\ll L_{0},L_{B}$, we
found $\delta _{x}\simeq 2.12L_{F};$ $\delta _{y}\simeq 1.73L_{F}.$ The
localization parameters obtained from the minimization of $\varepsilon
^{(10)}$ are $\sqrt{2}$ times smaller. If $B=0,$ the analytical solution of
Eqs. (\ref{3.15}) can be found in the limit $L_{0}\ll L_{F}$: $\delta
_{x}\simeq \sqrt{3}L_{F};$ $\delta _{y}\simeq L_{0}.$ Finally for high $B$ ($%
\omega _{c}\gg \omega _{0}$) one obtains $\delta _{x}\simeq $ $\delta
_{y}\simeq L_{B}$.

We now discuss the energy gap $\varepsilon ^{(10)}-\varepsilon ^{(0)}$
between the ground and excited states of the electron trapped in the dimple.
The numerical results of the spectroscopic transition frequency $\omega
_{10-0}=\varepsilon ^{(10)}-\varepsilon ^{(0)}/\hbar $ are presented in Fig.
3 as a function of the holding field for zero magnetic field and two
substrates $^{3}$He and $^{4}$He. In the range $1<E_{\perp }^{*}<3$ kV/cm,
the curves can be described by the analytical expression $\omega
_{10-0}\simeq (eE_{\perp }^{*})^{2}/2\pi \sigma \hbar $, and changes from $%
9.8\times 10^{8}$ s$^{-1}$ to $8\times 10^{9}$ s$^{-1}$ for the polaron over 
$^{4}$He which corresponds to increasing electron energy from $7.5\times
10^{-3}$ K to $6.2\times 10^{-2}$ K. This increase is significantly smaller
than $\left| E_{b}\right| $ calculated by Eq. (\ref{3.8}), and the electron
transition from ground to excited states does not destroy the polaron state.
It is interesting to note that $W_{10}-W_{0}\simeq [\varepsilon
^{(10)}-\varepsilon ^{(0)}]/2$ in such conditions is significantly smaller
than $|E_{b}|$ and that also $\omega _{10-0}\ll \omega _{0}$, which
characteristic values vary from $5.9\times 10^{10}$ s$^{-1}$ to $1.0\times
10^{11}$ s$^{-1}$ when $E_{\perp }^{*}$ increases from $1$ to $3$ kV/cm. One
should emphasize that the spectroscopic frequencies $\omega _{01-0}$ for the
transition from the ground to the excited state $\left| 0,1\right\rangle $
are significantly higher than $\omega _{10-1}$ and are the same as $\omega
_{0}.$ It means that after such a spectroscopic transition the electron
energy increases from $0.4$ K to $0.7$ K when $E_{\perp }^{*}$ lies in the
range of ($1-3$) kV/cm. These energies are significantly higher than $%
|E_{b}| $ and this transition should destroy the polaron state. For this
reason only the state $\left| 1,0\right\rangle $ can be considered as the
excited polaron state. The surface deformation for this state is given by

\begin{equation}
\xi _{10}(0,0)\simeq -\frac{eE_{\perp }^{*}}{2\pi \sigma }\left[ \ln \frac{4%
}{\sqrt{\gamma }k_{c}(\delta _{x}+\delta _{y})}-\frac{\delta _{x}}{\delta
_{x}+\delta _{y}}\right]  \label{3.16}
\end{equation}
which follows from Eq. (\ref{3.13}) in the limit of $k_{c}\sqrt{\delta
_{x}^{2}+\delta _{y}^{2}}\ll 1.$ The absolute values of $\xi _{10}(0,0)$ are
smaller than those of $\xi _{00}(0,0\dot{)}$. For $E_{\perp }^{*}\ =3$
kV/cm, $\xi _{10}(0,0)\simeq -1.67\times 10^{-8}$ cm for $B=0$.

The problem of recovering the symmetric case from the previous results in
the case of excites states is more complicated than for the ground state. If 
$\omega _{0}=0,$ the excited state can be described by the trial wave
function\cite{cheng78,far-peet97} 
\begin{equation}
\psi _{1}(r,\varphi )=\left( \frac{1}{\sqrt{\pi }\Delta ^{2}}\right) r\exp
(-r^{2}/2\Delta ^{2})e^{il_{z}\varphi }\ \ l_{z}=\pm 1,  \label{3.17}
\end{equation}
which is eigenfunction of the angular moment $\widehat{L}_{z}$ corresponding
to the excited state of the 2D harmonic oscillator in polar coordinates and $%
\Delta $ is the single localization parameter. However $\psi _{10}(x,y)$ is
not an eigenfunction of $\widehat{L}_{z}$ and hence cannot be reduced to $%
\psi _{1}(r,\varphi )$. For this reason the formal dependence of the excited
polaron state on two localization lengths $\delta _{x}$ and $\delta _{y}$
must be kept for $\omega _{0}=0$. In particular $\delta _{x}\neq \delta _{y}$
even at very high $E_{\perp }^{*}$ ($L_{F}\ll L_{0}$) where the effects of
the potential confinement are negligible and the properties of the symmetric
excited polaron state should be reproduced. For $\delta _{x}\simeq 2.12L_{F}$%
, and $\delta _{y}\simeq $ $1.73L_{F}$ ($B=0$), the symmetric polaron energy
in the excited state of symmetric can be written as 
\[
W_{1}\simeq -\frac{(eE_{\perp }^{*})^{2}}{4\pi \sigma }\left[ \ln \frac{0.550%
}{k_{c}L_{F}}-0.837\right] {.}
\]
Only at high magnetic fields ($L_{B}\ll L_{B}$ and $\omega _{c}\gg \omega
_{0}$) one has $\delta _{x}\simeq \delta _{y}$, which means that the
influence of $B$ on the electron localization is dominant over confinement
effects in the limit when the scale of electron localization is determined
by the magnetic length $L_{B}$.

It is interesting to know about the structure of the dimple at $r=0$ in the
limit of $\omega _{0}=0$. To make the problem more clear, we consider the
second-order derivative $\left[ \xi _{rr}^{^{\prime \prime }}\right]
_{10}(0) $ in the point of $r=0$ starting from wave function given by Eq. (%
\ref{3.11}) with $\delta _{x}\simeq 2.12L_{F}$ and $\delta _{y}\simeq
1.73L_{F}$. However for the trial wave function $\psi _{10}(x,y)$, $\left[
\xi _{rr}^{\prime \prime }\right] _{10}$ depends explicitly on $\varphi $
after conversion to polar coordinates. This dependence has no physical sense
for the symmetric polaron state and results from the inadequacy of $\psi
_{10}$ to describe correctly the excited state in the limit of $\omega
_{0}=0 $. Since in the symmetric case, the choice of the coordinate system
is arbitrary, one can consider the angle-averaged value of the second-order
derivative. One can easily show that $\langle \left[ \xi _{rr}^{^{\prime
\prime }}\right] _{10}(0)\rangle $ and $\langle \left[ \xi _{rr}^{^{\prime
\prime \prime }}\right] _{10}(0)\rangle $ are almost zero in the limit of $%
k_{c}^{2}(\delta _{x}^{2}+\delta _{y}^{2})\ll 1$. Considering higher-order
terms, one finds $\langle \left[ \xi _{rr}^{IV}\right] _{10}(0)\rangle
\simeq 3eE_{\perp }^{*}\ /2\pi \sigma \delta _{x}^{3}\delta _{y}$ which
suggests a minimum in the centre of the isotropic dimple. This estimative
agrees with the results of numerical calculations of Ref. \cite{far-peet97}.
We have observed that at large distances the function $\xi _{1}(r)$
decreases exponentially in the same manner as $\xi _{0}(r)$.

We also can extend our formalism to consider the polaron impurity states
treated in Ref. \cite{far-peet97}. The authors calculated the energy gap
between the ground and excited electron states over a helium film when a
localization potential is considered due to a positive impurity charge $Ze$
located on the top of the substrate supporting the film with thickness $d$.
It is easy to show that the impurity potential gives a correction $Ze/d^{2}$
to $E_{\perp }^{*}$ and an additional parabolic term $m\varpi
^{2}(x^{2}+y^{2})/2$ with $\varpi ^{2}=Ze^{2}/md^{3}$, which turns the
problem similar to that considered in the present work. Making the necessary
adjustments of the expressions of $W^{(0)}$ [$\varepsilon ^{(0)}$] and
taking the same parameters $d,k_{c}$ and $Z$ as in Ref. \cite{far-peet97} we
obtain $\varepsilon ^{(10)}-\varepsilon ^{(0)}\simeq 0.60$ meV which should
be compared with the value $0.445$ meV obtained by Farias and Peeters in a
fully numerical calculation.\cite{far-peet97}

\section{POLARON\ TRANSPORT}

We now investigate the transport properties of the Q1D polaron. When a
driving electric field $E_{\Vert }$ is applied along the plane $xy$, the
surface deformation moves together with the trapped electron inducing a
field of hydrodynamic velocities in the liquid, which is accompanied by
energy dissipation, and leads to a finite value of the polaron mobility.\cite
{sh-mon73,mon75,review,marq-stud89} This approach is valid only in the
strong coupling limit (high $E_{\bot }^{*}$) where the self-trapped state is
energetically favored in comparison with the weak coupling limit where the
electron is simply scattered by ripplons at $T<1$ K.\cite{sok-hai-stud95} In
order to evaluate the polaron mobility, we employ the energy balance
equation $eE_{\Vert }v_{0}=d\rho _{E}/dt,$ where $v_{0}$ is the liquid
velocity at infinity, and $\rho _{E}$ is the energy density dissipated. The
function $d\rho _{E}/dt$ is obtained in a straightforward way by finding the
normal velocity field induced by the polaron from the solution of the
Navier-Stokes equation. The calculation procedure in the asymmetric case is
similar to that used in Refs. \cite{sh-mon73,mon75} for the symmetric
polaron and are based in the same system of equations and boundary
conditions. However the asymmetry of the surface deformation in $x$ and $y$
directions should be taken into account and 2D Fourier transforms like Eqs. (%
\ref{2.7}) should be employed. If the driving field is applied along the $x$ 
$(y)$ direction, the mobility is given by 
\begin{equation}
\mu _{x(y)}=\frac{e}{2\eta S}\left[ \sum_{{\bf k}}kk_{x(y)}^{2}|\xi _{{\bf k}%
}|^{2}\right] ^{-1},  \label{4.1}
\end{equation}
where $\eta $ is the helium viscosity. The summation in Eq. (\ref{4.1}) can
be performed analytically and the polaron mobility in the case of $E_{\Vert
} $ along the $x$ direction can be written in terms of the complete elliptic
integrals {\bf K}$(t)$ and {\bf E}$(t)$ as

\begin{eqnarray}
\mu _{x} &=&\frac{\pi ^{3/2}\sigma ^{2}}{\sqrt{2}e(E_{\perp }^{*})^{2}\eta }%
{\LARGE \{}\frac{(\ell _{x}^{2}-\ell _{y}^{2})\Theta (\ell _{x}-\ell _{y})}{%
\ell _{x}\left[ \text{{\bf E}}\left( \sqrt{1-\ell _{y}^{2}/\ell _{x}^{2}}%
\right) -(\ell _{y}/\ell _{x})^{2}\text{{\bf K}}\left( \sqrt{1-\ell
_{y}^{2}/\ell _{x}^{2}}\right) \right] }  \label{4.2} \\
&&+\frac{(\ell _{y}^{2}-\ell _{x}^{2})\Theta (\ell _{y}-\ell _{x})}{\ell
_{y}\left[ \text{{\bf E}}\left( \sqrt{1-\ell _{x}^{2}/\ell _{y}^{2}}\right) -%
\text{{\bf K}}\left( \sqrt{1-\ell _{x}^{2}/\ell _{y}^{2}}\right) \right] }%
{\LARGE \}},  \nonumber
\end{eqnarray}
Here $\Theta (t)$ is the step function. The general Eq. (\ref{4.2}) is quite
simplified in the following limiting cases:

\begin{equation}
\mu _{x}\simeq \frac{\pi ^{3/2}\sigma ^{2}\ell _{x}}{\sqrt{2}e(E_{\perp
}^{*})^{2}\eta }\ \text{ if }\ell _{x}\gg \ell _{y}  \label{4.3}
\end{equation}
and 
\begin{equation}
\mu _{x}\simeq \frac{\pi ^{3/2}\sigma ^{2}\ell _{y}}{\sqrt{2}e(E_{\perp
}^{*})^{2}\eta \ln (\ell _{y}/\ell _{x})}\text{ \ if\ }\ell _{x}\ll \ell
_{y}.  \label{4.4}
\end{equation}
The polaron mobility along the $y$ direction, $\mu _{y}$, is the same as for 
$\mu _{x}$ but the index $``x"$ must be replaced by $``y"$ in Eqs. (\ref{4.2}%
)-(\ref{4.4}) and vice versa. If $\ell _{x}=\ell _{y}=\ell ,$ both $\mu _{x}$
and $\mu _{y}$ reproduce the mobility of the symmetric polaron given by\cite
{sh-mon73,marq-stud89}

\[
\mu =\frac{\sqrt{8\pi }\sigma ^{2}\ell }{e(E_{\perp }^{*})^{2}\eta }. 
\]
As is was seen in Sec. II, the limiting case $\ell _{x}=\ell _{y}=\ell
\simeq \sqrt{2}L_{F}$ can be reached in the limit of extremely high holding
fields $E_{\perp }^{*}\gg 4$ kV/cm which makes this limit experimentally
inaccessible . For $1<E_{\perp }^{*}<3$ kV/cm, we have $\ell _{x}\simeq
L_{F}\gg \ell _{y}\simeq L_{0}.$ In such a condition the longitudinal the
polaron mobility $\mu _{l}$ (along the $x$ direction of channel) is given by
Eq. (\ref{4.3}). The transversal mobility $\mu _{t}$ (along the $y$
direction across the channel) is defined by Eq. (\ref{4.4}). As it follows
from Eqs. (\ref{4.3}) and (\ref{4.4}), $\mu _{l}/\mu _{t}\sim \ln (\ell
_{x}/\ell _{y})$ if $\ell _{x}\gg \ell _{y}$.

The holding field dependences of $\mu _{l}$ and $\mu _{t}$ are depicted in
Fig. 5. The magnetic field influences strongly the mobility because it
decreases the localization parameters as it can be seen in Fig. 5 where
polaron mobilities are plotted as functions of $E_{\perp }^{*}$ for some
values of $B$. Since $\mu \sim \sigma ^{2}$ and $\sigma _{4}\simeq 2\sigma
_{3},$ the mobilities for the polaron over $^{4}$He are higher than that for 
$^{3}$He as the substrate.

\section{CONCLUDING REMARKS}

In this work, we have discussed the possibility for the formation of the
polaron in the context of anisotropic liquid helium surface. In particular,
we have evaluated the energetics and transport properties of the polaron in
a Q1D channel. We have considered the properties of both ground and excited
states of the asymmetric polaron. The localization lengths and the ground
and excited state energies have been determined as functions of the holding
electric field within the hydrodynamical model and using a variational
approach. We have obtained the frequency of spectroscopic transitions
between the ground and first levels in the asymmetric polaron states. The
transport properties of the polaron along the corrugated helium surface have
been also studied.

We think that the measurement of the frequency of spectroscopic transition $%
\omega _{10-0}$ between ground and excited states (see Fig. 4) should be
favorable in experimental attempts to observe the polaron state in Q1D
channel over liquid helium surface. Indeed, our estimates indicate that $%
\omega _{10-0}$ differs significantly of $\omega _{0}$ of spectroscopic
transitions between subbands of free electron state in Q1D channel which
allows to separate the experimental signals for these two different types of
spectroscopy transitions.

Another possibility is measuring directly the mobility as in Ref.\cite{tress}%
. Our results point out that the mobility of the Q1D polaron is $40\%$
smaller than the mobility of the 2D polaron for the same holding field in
the range $E_{\perp }^{*}<3$ kV/cm. Our value of the Q1D polaron mobility $%
\sim 10^{6}$ cm$^{2}$/Vs for $T=0.1$ K and $E_{\perp }^{*}=3$ kV/cm, is
three orders of magnitude smaller than the mobility of 2D electrons over a
flat helium surface and that for electrons moving freely along the channel%
\cite{sok-hai-stud95}, which allow us to easily distinguish them easily.

While the above estimates show that our proposed mechanism for the polaron
formation in Q1D systems on helium should be compatible with accessible
experiments, we add other comments to corroborate or refute our predictions.
Of particular interest is the possibility to use liquid $^{3}$He instead $%
^{4}$He as the substrate for the confined Q1D electron system.\cite{kono95}
Recall that for $^{3}$He, whose surface tension coefficient is more than
twice smaller than that of $^{4}$He, the binding energy $|E_{b}|$ can be
substantially larger. Also the spectroscopic transition frequencies are
twice larger for $^{3}$He in comparison with $^{4}$He. Furthermore, the
description of the polaron in terms of a hydrodynamic viscous model may be
doubtful for pure $^{4}$He for temperatures below $1$ K. However, we must
emphasize that experimental results\cite{tress} have been successfully
interpreted by theoretical calculations from the hydrodynamic approach for $%
T<1$ K, even though this interpretation has been seriously questioned.\cite
{dahm} Furthermore the hydrodynamic approach of viscous phenomena is well
satisfactory in the case of $^{3}$He for $T$ down to $\sim 0.1$ K. Other
attractive substrate should be also the low $T$ mixture of $^{3}$He and $%
^{4} $He.\cite{hallock} Recent experiments on Q1D-channels have provided a
higher effective holding field as for electrons on helium films which could
greatly enlarging the absolute value of the binding energy $E_{b}.$\cite
{yayama}

The electron states have been described in the present work in the
one-electron approximation. The validity of the results in the case of
finite electron densities is limited by the condition that the scale of
electron localization is small in comparison with the mean interelectronic
distance $a\simeq n_{l}^{-1}$ along the channel axis where $n_{l}$ is linear
electron density. For $B=0$ and $E_{\perp }<3$ kV/cm where $\ell _{x}\simeq
L_{F}\gg \ell _{y}\simeq L_{0}$ the characteristic values of $\ell _{x}\sim
10^{-5}$ cm which conditions $n_{l}\ll 10^{5}$ cm$^{-1}$. At higher electron
densities, correlation effects could strongly influence the polaron
properties and the applicability of our theory developed for the low-density
limit becomes doubtful. For low electron densities, the experimental
difficulties may appear in providing the measurements of experimental signal
of the response of electron system with low level. This signal is
proportional to the electron density which is especially restricted if we
have a single Q1D channel. One possibility should be to use weakly coupled $%
N_{ch}$ multi-channels\cite{grating,gate} like in multiple quantum wells in
semiconductor heterostructures. If the average distance between the channels 
$b\gg a$ one can disregard the correlation between electrons in different
channels and the total response should be enhanced by the factor $N_{ch}$.
This turns more appropriate the conditions for experimental investigation of
Q1D electron systems over liquid helium in the low-density limit.

One hope that the asymmetric polaron can be detected at low temperatures,
probably at $T<0.1$ K, and modern experimental methods for achieving low and
ultralow temperatures would offer the possibility to observe the polaron in
the Q1D channel on the liquid helium surface.

\section{ACKNOWLEDGMENT}

This work was partially sponsored by the Funda\c{c}\~{a}o de Amparo \`{a}
Pesquisa do Estado de S\~{a}o Paulo (FAPESP) and the Conselho Nacional de
Desenvolvimento Cientifico e Tecnol\'{o}gico (CNPq), Brazil.

\medskip

\begin{center}
{\bf FIGURE CAPTIONS}
\end{center}

\medskip

Fig. 1. Localization parameters $\ell _{x}$ (a) and $\ell _{y}$ (b) in units
of $L_{F}=\left( 2\pi \hbar ^{2}\sigma /m\right) ^{1/2}/(eE_{\perp }^{*})$
for the polaron ground-state on the surface of ${^{4}}$He as a function of
the holding electric field $E_{\perp }^{*}$ for some values of the magnetic
field.\smallskip

Fig. 2. Total energy of the polaron ground-state as a function of the
holding electric field on the surface of ${^{4}}$He.\smallskip

Fig. 3. Frequency of spectroscopic transition from the ground to excited
electronic states in the asymmetric dimple as a function of the holding
electric field for ${^{4}}$He and ${^{3}}$He liquid surfaces.\smallskip

Fig. 4. Longitudinal $(\mu _{l}$) and transversal $(\mu _{t})$ polaron
mobilities over $^{4}$He as a function of the holding electric field
calculated using by Eq. (\ref{4.2}) (solid line) and by the approximate
expression [Eqs. (\ref{4.3}) and (\ref{4.4})] (dashed line) for zero
magnetic field.\smallskip

Fig. 5. Longitudinal polaron mobility versus the holding electric field for
some values of the magnetic field.

\end{document}